\documentclass[10pt,letterpaper]{article}

\usepackage{cogsci}
\cogscifinalcopy 
\usepackage{graphicx} 

\usepackage[
  style=apa,
  natbib=true,
  annotation=false,
]{biblatex}
\usepackage{float} 

\title{Analyzing Human Heuristics and Strategies in Everyday Decision-Making Conversations for Conversational AI Design}

\author[]{
  \mbox{Sora Kang (sorakang@snu.ac.kr)},
  \mbox{Soyun Jeon},
  \mbox{Jinsu Eun},
  \mbox{Kwangwon Lee} \\
  \mbox{Chaerin Song}, 
  \mbox{Minyoung Joo},
  \mbox{Joonhwan Lee}
}

\affil[]{Human-Computer Interaction+Design Lab, Seoul National University, Seoul, Republic of Korea}

\begin{document}

\maketitle
\begin{abstract}
Conversational AI increasingly supports everyday decision-making, yet most systems rely on data-centric reasoning rather than the heuristic and interactional strategies people use in natural conversation. To ground design in actual human practice, we analyze 955 real-world Korean conversations (15,476 utterances) involving food and travel decisions, applying a decision-making codebook through an LLM-assisted coding pipeline. Our findings reveal that people prioritize satisficing over optimization, relying heavily on internal knowledge and interactional strategies to manage cognitive load. Critically, we identify a frequency–efficiency mismatch: the most prevalent heuristics sustain conversational flow during exploration, whereas infrequent, rule-based strategies are highly effective at driving resolution during exploitation. By mapping how these patterns transfer across the spectrum of human-AI interaction, this work provides empirical grounding consistent with cognitive theories of decision-making and offers design implications that align AI systems with human heuristic processes.


\end{abstract}

\section{Introduction}
Recent advances in large language models (LLMs) have positioned conversational AI as an increasingly capable interaction partner, not merely a tool for information retrieval. Yet most contemporary conversational systems still operate primarily as answer providers: they respond to well-formed questions, summarize relevant information, and produce recommendations given explicit constraints. This framing fits tasks with clear correctness criteria (e.g., factual queries or constrained lookups). In contrast, much of everyday life consists of decisions without a single objective “right” answer—what to eat tonight, where to travel next, or which hobby to start—where choices are shaped by personal values, affect, context, and limited time.

Decision-making in such settings is classically explained through bounded rationality and satisficing (i.e., seeking an option that is ``good enough'' rather than globally optimal) \citep{simon1955behavioral, simon1956rational}. Under bounded resources, people rely on heuristics---fast, experience-based shortcuts---to reduce cognitive effort and reach closure. These heuristics can be adaptive and efficient, but they also introduce systematic distortions and biases. For a conversational AI meant to support everyday decision-making, this creates a central design tension: effective support should feel lightweight and conversation-friendly (as in human dialogue), while also helping users structure reasoning, reduce bias, and maintain agency. Yet we still lack clear empirical answers to a key question: How do humans actually talk their way toward everyday decisions, and what heuristics and interactional strategies structure this process?

Moreover, conversational decision-making encompasses a range of participant structures. Two friends may co-decide where to eat, one person may recommend options while the other chooses, or a knowledgeable party may guide an uncertain one toward a decision. These configurations---co-decider, recommender, advisor---reflect the fact that language use in decision-making is fundamentally a coordinated activity whose structure varies with the roles, stakes, and relationships of the participants \citep{clark1996using}. Understanding how cognitive heuristics and interactional strategies operate across these varied structures is essential not only for theories of human decision-making but also for designing conversational AI systems, which may occupy any of these roles.
 
To address this gap, we propose a human-grounded approach that treats everyday decision-making as a conversational process rather than a one-shot recommendation problem. We analyze real human–human decision dialogues to identify recurring strategies and heuristics as they appear in conversation, and we compile these patterns into a reusable codebook. This paper investigates the following research questions:

\begin{itemize}
    \item RQ1. How are decision-making strategies and heuristics manifested and distributed in everyday human–human decision-making conversations?
    \item RQ2. Which conversational strategies and heuristics are associated with whether a decision is ultimately reached?
\end{itemize}

Our contributions are: (1) a conversation-grounded codebook of decision-making heuristics and strategies; (2) an LLM-assisted coding pipeline for large-scale analysis; and (3) an empirical characterization of how conversational strategies relate to decision outcomes, with implications for conversational AI design.

\section{Related Work}
\subsection{Decision-Making Theories: Heuristics and Biases}
Human decision-making is defined by cognitive and environmental constraints, contrasting the rational agent model with Simon’s theory of bounded rationality \citep{simon1955behavioral}. Individuals navigate these limits through satisficing—selecting "good enough" options rather than exhaustive optimization \citep{simon1955behavioral, simon1956rational}. Under bounded resources, people rely on heuristics: fast, intuitive shortcuts that reduce cognitive effort \citep{tversky1974judgment}. These include judgment heuristics such as availability \citep{tversky1973availability}, representativeness \citep{tversky1974judgment}, affect \citep{slovic2002affect}, and anchoring \citep{tversky1974judgment}. Furthermore, choices are sensitive to framing effects and loss aversion, which skew decisions based on presentation \citep{tversky1981framing, kahneman1979prospect}. Decision-making is also a social process influenced by heuristics like consistency, reciprocity, and social proof \citep{cialdini2001influence}. For selection tasks, adaptive agents utilize choice heuristics—such as lexicographic rules or elimination by aspects—flexibly switching between them based on task complexity \citep{gigerenzer1999simple, payne1993adaptive}. While functional, these processes can result in systematic biases \citep{kahneman2011thinking}.

\subsection{Conversational AI and Decision Support}
AI-assisted decision-making aims for complementary performance between human and machine \citep{bansal2021does, lai2023towards, reverberi2022experimental, zhang2024beyond}. However, recommendation-centric systems often ignore the underlying reasoning process, leading to inappropriate reliance where users either over-rely on or disregard suggestions \citep{lai2023towards, ma2023who}. While AI explanations are intended to foster trust, they can paradoxically increase over-reliance by reducing active cognitive engagement \citep{bucinca2021trust, miller2023explainable}. Recent research thus advocates for AI as a cognitive assistant that encourages independent judgment \citep{ma2025towards, park2023thinking}. Systems like Extend AI and frameworks focusing on causal rather than probabilistic reasoning aim to align AI with human cognition and mitigate over-reliance \citep{miller2019explanation, reber2025aihelpme}. While analyses of user--chatbot interaction logs have informed such designs \citep{zhang2021forward, li2020conversation}, these logs often reflect existing system limitations rather than natural interaction; human--human dialogues reveal richer relational cues and patterns for designing more empathic and effective agents \citep{salman2021analysis}. We build on this by empirically characterizing naturalistic decision-making talk to ground future AI that aligns with human heuristics.

\section{Method}
This paper presents a conversation-analytic study of everyday decision-making interactions. We analyze real-world conversations to uncover the strategies individuals use in decision-making and to characterize how heuristics surface in naturalistic dialogue.

\subsection{Human Conversation Dataset}

To examine how individuals interact and address the inherent challenges of decision-making, we employed the Subject-Specific Text Data for Everyday Conversation dataset from AI-Hub \citep{aihub_everyday_conv_2021}, a Korean national platform providing open AI resources. The dataset consists of free-form conversations in Korean, spanning 20 different subjects. For this study, we focused on two domains—Food \& Drink (5,176 conversations) and Travel (5,550 conversations)—as they frequently involve collaborative decision-making tasks, such as choosing meals or selecting travel destinations. From these subsets, we filtered for conversations containing explicit decision-making sequences, excluding those lacking a clear decision goal (e.g., simple reporting of past events) or containing insufficient interactional turns. After this filtering, a set of 955 conversations (15,476 utterances) remained, which served as the basis for both codebook validation and analysis.

\subsection{Conversation Codebook}

\begin{table*}[t]
\centering
\caption{The Decision-Making Conversation Codebook. The framework consists of conversation-level outcomes and methods, utterance-level interaction structures, and cognitive heuristics.}
\label{tab:final_codebook_fixed}
\small
\renewcommand{\arraystretch}{1.3}
\begin{tabular}{l p{0.75\textwidth}}
\hline
\textbf{Category} & \textbf{Definitions and Indicators} \\ \hline

\multicolumn{2}{l}{\textit{\textbf{I. Conversation Level Analysis}}} \\
\hspace{3mm}\textbf{Outcome} & \textbf{Decision Reached:} Interaction concludes with a mutual decision. / \textbf{No Decision:} Unresolved. \\ 
\hspace{3mm}\textbf{Method} & \textbf{Satisficing:} Stopping the search once a ``good enough'' option is found. Prioritizes sufficiency. \newline
\textbf{Maximizing:} Considering all variables and alternatives to find the ``optimal'' choice. \\ \hline

\multicolumn{2}{l}{\textit{\textbf{II. Utterance Level Analysis (Structure)}}} \\
\hspace{3mm}Info. Collection & \textbf{Source:} \textit{External} (Objective facts, reviews) vs. \textit{Internal} (User preferences, situations). \newline
\textbf{Type:} Covers \textit{Schedule, Cost, Activity, Place, Time, Reputation, Other.} \\ 

\hspace{3mm}Suggestion & \textbf{Suggestion:} Proposing clear options or recommendations. \newline
\textit{-- Context-based:} Grounded in previous dialogue context. \newline
\textit{-- Explanation-based:} Accompanied by reasoning or justification. \\ 

\hspace{3mm}Response & \textbf{Explicit:} Agreement (Positive preference) vs. Rejection (Negative preference). \newline
\textbf{Implicit Agreement:} Positive cues or feedback without explicit confirmation. \newline
\textbf{Implicit Rejection:} \newline
-- \textit{Alternative (Type 1):} Indirect refusal by suggesting a different option. \newline
-- \textit{Avoidance (Type 2):} Indirect refusal by evading the decision. \newline
-- \textit{Reasoning (Type 3):} Indirect refusal by providing reasons for dislike. \newline
\textbf{Other:} \textit{Indifference (NM)} (``No matter'' but hidden preference), \textit{Suspension} (Deferring decision). \\ 

\hspace{3mm}Strategies & \textbf{Choice Overload Mitigation (COM):} Narrowing the option set to reduce cognitive load. \newline
\textbf{Open-Ended Question (OEQ):} Broadening the scope to invite free-form opinions. \\ \hline

\multicolumn{2}{l}{\textit{\textbf{III. Cognitive Heuristics Analysis}}} \\
\hspace{3mm}Judgment & \textit{Availability} (Recall ease), \textit{Representativeness} (Similarity), \textit{Anchoring} (Reference point), \textit{Affect} (Emotion), \textit{Framing} (Gain/Loss presentation). \\ 
\hspace{3mm}Social Influence & \textit{Consistency}, \textit{Reciprocity}, \textit{Scarcity}, \textit{Social Proof}, \textit{Likability}, \textit{Authority}. \\
\hspace{3mm}Choice Rules & \textit{Lexicographic} (Most important attribute), \textit{Attribute Elimination} (Sequential rejection), \textit{Additive Difference}, \textit{Conjunctive}/\textit{Disjunctive} (Thresholds). \\ \hline
\end{tabular}
\end{table*}

The codebook was developed through an iterative process to adapt theoretical concepts to systematic conversation analysis. The initial schema was grounded in established behavioral economics and decision theory, drawing on Simon's satisficing framework \citep{simon1955behavioral}, Tversky and Kahneman's heuristics \citep{tversky1974judgment}, and Slovic's affect heuristic \citep{slovic2002affect}. We conducted a pilot coding process on a random sample of 50 conversations from the initial AI-Hub pool (distinct from the final analysis set), followed by reconciliation sessions among the research team to review discrepancies, identify ambiguities, and clarify category definitions. Through this consensus-building process, we refined the codebook iteratively until the coders reached a shared understanding.
 
Through this refinement, the codebook evolved to capture specific conversational nuances. New dimensions were added, such as distinguishing the \textit{source} of information (internal preferences vs.\ external facts) and specifying the \textit{type} of information exchanged. The Suggestion and Response codes were expanded to include finer-grained response types (e.g., implicit agreement or soft deferral), moving beyond simple binary classifications. Overlapping or unclear categories were consolidated or removed to improve coding consistency.
 
\paragraph{Conversation-Level Codes.}
We coded the conversations on two dimensions. \textit{Outcome} captures whether a decision was reached or if the dialogue ended unresolved. \textit{Decision Method} identifies the strategy used: \textit{Satisficing} refers to selecting a ``good enough'' option without extensive search, whereas \textit{Maximizing} involves seeking the optimal choice by comparing multiple variables.
 
\subsubsection{Utterance-Level Codes}
At the utterance level, our coding framework is organized into five dimensions.
 
\textit{Information Collection} categorizes utterances where participants request or provide decision-relevant information from external sources (e.g., online reviews, third-party opinions) or internal knowledge (e.g., personal preferences, prior experiences), covering types such as price, time, or logistics. It applies whenever information necessary for a decision or for proposing an option is revealed; simple exclamations or reactions are not coded.
 
\textit{Suggestion} includes utterances proposing specific options or recommendations. Suggestions may be grounded in previous dialogue context or accompanied by explanations justifying why a particular option should be considered.
 
\textit{Response} identifies utterances in which a speaker reacts to a previous suggestion or piece of information. Responses may express agreement or rejection directly, hint at acceptance or rejection indirectly, indicate indifference, or signal that the decision should be put on hold. This dimension captures the dynamics of acceptance, negotiation, resistance, and deferral in decision-making conversations.
 
\textit{Choice Overload Mitigation (COM)} refers to utterances that reduce cognitive load by narrowing available alternatives, often presenting a limited set of options (e.g., ``How about pizza or pasta?'') rather than an open-ended array.
 
\textit{Open-Ended Question} identifies utterances that invite the other party to respond freely by sharing opinions, preferences, or alternative suggestions. Unlike COM, open-ended questions deliberately broaden the conversation scope (e.g., ``What do you feel like eating?''), creating opportunities for participants to express unarticulated preferences.

\subsubsection{Heuristics}
The heuristic level comprises three dimensions (Table~\ref{tab:final_codebook_fixed}). \textit{Judgment Heuristics} code intuitive shortcuts in evaluating options or probabilities: availability (recall ease), representativeness (similarity to a familiar category), anchoring (initial reference point), affect (immediate emotional reactions), and framing (gain/loss presentation). \textit{Social Influence Heuristics} code appeals to social cues: consistency, reciprocity, scarcity, social proof, likability, and authority. \textit{Choice Heuristics} code simplified selection rules: lexicographic (best on the most important attribute), attribute elimination (sequential exclusion), additive difference (weighted comparison), and conjunctive/disjunctive (minimum thresholds).
 
\subsection{LLM-Assisted Coding Procedure}
Following the development of the codebook through human coding, we applied it to the full dataset using a human-in-the-loop validation pipeline to ensure reliability. We utilized the GPT-4 model for automatic coding. To bridge the gap between human nuance and machine processing, we implemented an iterative refinement process. Initial trials revealed discrepancies, particularly in subtle categories such as implicit rejection and choice overload mitigation. We addressed this by incorporating concrete examples derived from the pilot study directly into the system prompt to guide the model's reasoning, effectively utilizing few-shot prompting. Furthermore, we conducted manual reviews on random batches of 20 conversations after each prompt adjustment. We compared the output against human consensus and refined the definitions until the model demonstrated consistent alignment with human annotators. We proceeded to full-scale coding of the 955 conversations only after three consecutive batches showed no major deviations from human judgment.

\section{Results}
\subsection{Decision Outcomes and Methods}
 
As summarized in Table \ref{tab:descriptive_stats}, among the 955 conversations, 693 (72.6\%) resulted in a decision. Across all conversations, satisficing (41.9\%) was the most frequent strategy, significantly exceeding maximizing (15.7\%) ($\chi^2 = 113.6, p < .001$), suggesting a general preference for cognitive economy over exhaustive comparison.

\subsection{Utterance-Level Patterns}
 
\subsubsection{Information Collection}
Information collection was the most frequent code (45.2\%). A Chi-square test reveals a significant reliance on internal knowledge (55.1\%) over external information (14.4\%) ($\chi^2 = 1668.5, p < .001$), indicating that everyday decision-making operates primarily through memory retrieval rather than active information discovery.
 
\subsubsection{Suggestions and Responses}
Suggestions occurred in 1,992 utterances, with 63.1\% relying solely on context without justification. In responses ($N=5,785$), explicit agreement was common (33.2\%), whereas explicit rejection was rare (1.9\%). Instead, participants utilized indirect disagreement strategies (combined 38.1\%), such as implicit deferral or rejection.

\begin{table}[H]
\centering
\caption{Descriptive Statistics: Conversations and Utterances}
\label{tab:descriptive_stats}
\small
\begin{tabular}{lrr}
\hline
\textbf{Category} & \textbf{Count} & \textbf{Percentage} \\ \hline
\multicolumn{3}{l}{\textbf{Conversation Level} ($N=955$)} \\
\hspace{3mm}Decision Reached & 693 & 72.6\% \\
\hspace{3mm}No Decision & 262 & 27.4\% \\ 
\textit{\hspace{3mm}Method used (full sample):} & & \\
\hspace{6mm}Satisficing & 400 & 41.9\% \\
\hspace{6mm}Maximizing & 150 & 15.7\% \\ 
\hspace{6mm}Unclear/Mixed & 405 & 42.4\% \\ \hline
\multicolumn{3}{l}{\textbf{Utterance Level} ($N=15,476$)} \\
\textbf{1. Information Collection} & \textbf{6,993} & \textbf{45.2\%} \\
\hspace{3mm}Internal Knowledge & 3,853 & (55.1\%)* \\
\hspace{3mm}External Information & 1,006 & (14.4\%)* \\
\hspace{3mm}Source Unspecified/Other & 2,134 & (30.5\%)* \\
\textbf{2. Response} & \textbf{5,785} & \textbf{37.4\%} \\
\hspace{3mm}Explicit Agreement & 1,922 & (33.2\%)* \\
\hspace{3mm}Implicit Agreement & 1,303 & (22.5\%)* \\
\hspace{3mm}Implicit Rejection & 1,299 & (22.5\%)* \\
\hspace{3mm}Implicit Deferral/Suspension & 902 & (15.6\%)* \\
\hspace{3mm}\textit{Explicit Rejection} & \textit{112} & \textit{(1.9\%)*} \\
\hspace{3mm}\textit{Other (NM, N/A)} & \textit{247} & \textit{(4.3\%)*} \\
\textbf{3. Suggestion} & \textbf{1,992} & \textbf{12.9\%} \\
\textbf{4. Choice Overload Mitigation} & \textbf{1,433} & \textbf{9.3\%} \\ 
\textbf{5. Open-Ended Question} & \textbf{988} & \textbf{6.4\%} \\ \hline
\multicolumn{3}{r}{\textit{*Percentage within the specific category}} \\
\end{tabular}
\end{table}
 
\subsection{Conversational Strategies}
 
\subsubsection{Choice Overload Mitigation (COM)}
Choice overload mitigation appeared in 1,433 utterances (9.3\%). Crucially, almost half of all suggestions (48.8\%) were framed using this strategy, with speakers frequently narrowing the field by presenting binary contrasts (e.g., ``Pizza vs. Pasta'') rather than open lists.

\subsection{Heuristic Usage and Decision Success}

We analyzed heuristic usage ($N=3,416$) to examine the relationship between heuristic frequency and decision outcomes.

\subsubsection{Prevalence vs. Efficiency}
\begin{figure}[t]
    \centering
    \includegraphics[width=\columnwidth]{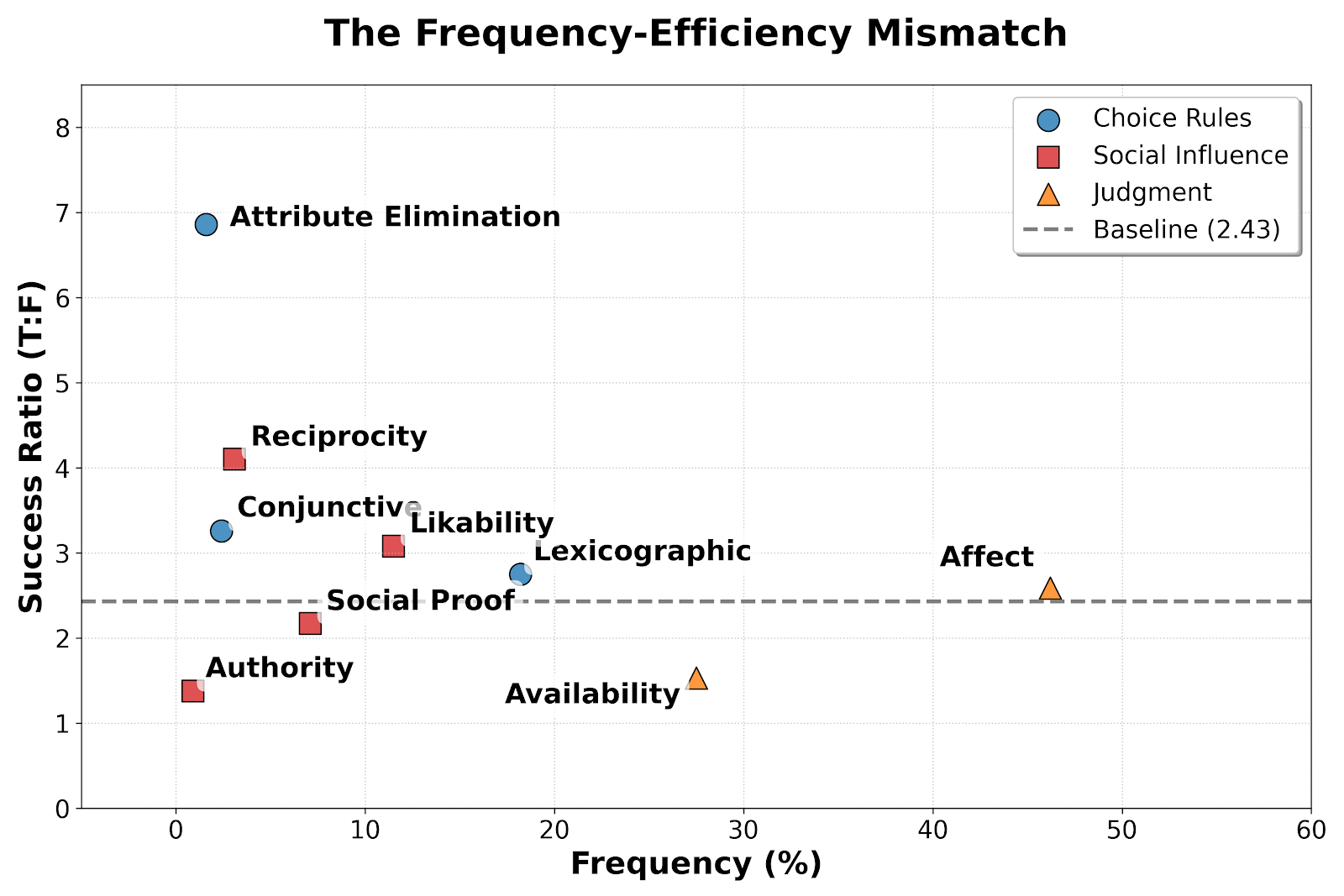} 
    \caption{Frequency--Efficiency Mismatch across heuristic types. High-frequency heuristics (e.g., Affect) show lower decision success ratios, whereas low-frequency rule-based strategies (e.g., Attribute Elimination) show the highest.}
    \label{fig:mismatch_scatter}
\end{figure}

As illustrated in Figure \ref{fig:mismatch_scatter} and detailed in Table \ref{tab:heuristics_ratio}, the data revealed a mismatch between the usage frequency of heuristics and their effectiveness in resolving decisions.
 
\begin{itemize}
    \item \textbf{High-Frequency, Low-Efficiency Heuristics:} Affect (46.2\%) and Availability (27.5\%) were the most frequently observed heuristics. However, their efficiency was relatively low; Availability showed a success ratio of 1.53:1.
     
    \item \textbf{Low-Frequency, High-Efficiency Heuristics:} Conversely, the most decisive heuristics were rare. Attribute Elimination (1.6\%) demonstrated the highest success ratio (6.86:1), significantly outperforming the baseline average ($p < .001$).
\end{itemize}

\subsubsection{Social Heuristics}
Strategies relying on external validation, such as Authority (1.38:1), were associated with lower decision success, whereas relational heuristics like Reciprocity (4.10:1) were highly effective.

\begin{table}[H]
\centering
\caption{Frequency and Success Ratio of Key Heuristics}
\label{tab:heuristics_ratio}
\small
\begin{tabular}{l c c}
\hline
\textbf{Heuristic} & \textbf{Freq. (\%)} & \textbf{Ratio (T:F)} \\ \hline
\multicolumn{3}{l}{\textit{Choice Rules}} \\ 
\hspace{3mm}Attribute Elimination & 1.6\% & 6.86 : 1 \\
\hspace{3mm}Conjunctive & 2.4\% & 3.26 : 1 \\
\hspace{3mm}Disjunctive & 10.3\% & 2.73 : 1 \\
\hspace{3mm}Lexicographic & 18.2\% & 2.75 : 1 \\
\hline
\multicolumn{3}{l}{\textit{Social Influence}} \\
\hspace{3mm}Reciprocity & 3.1\% & 4.10 : 1 \\
\hspace{3mm}Likability & 11.5\% & 3.08 : 1 \\
\hspace{3mm}Social Proof & 7.1\% & 2.17 : 1 \\
\hspace{3mm}Authority & 0.9\% & 1.38 : 1 \\ \hline
\multicolumn{3}{l}{\textit{Judgment}} \\
\hspace{3mm}Affect & 46.2\% & 2.59 : 1 \\
\hspace{3mm}Availability & 27.5\% & 1.53 : 1 \\ \hline
\textit{Baseline Average} & - & \textit{2.43 : 1} \\ \hline
\end{tabular}
\end{table}

\section{Discussion}
Everyday decision-making in conversation deviates significantly from classical optimization, operating instead as a satisficing process driven by heuristics and interactional strategies. In this section, we unpack the cognitive mechanisms underlying these conversational patterns, examine how this bounded rationality transfers to human--AI contexts, and outline design implications that align conversational AI with human cognitive processes.

\subsection{Cognitive Heuristics in Dialogue}

\subsubsection{Frequency--Efficiency Mismatch in Heuristics}

One of the central findings of this study is the inverse relationship between heuristic frequency and decision success (Figure~1). Affective heuristics were the most prevalent strategy but were not strongly associated with decision resolution, whereas rule-based strategies, such as Attribute Elimination, were rare yet exhibited the highest success ratio.

We propose that this mismatch reflects a functional division within conversational decision-making. High-frequency heuristics such as affect and availability serve a dual function: they operate as cognitive shortcuts for evaluating options, but they simultaneously perform a conversational function---maintaining dialogue flow, building shared context, and signaling engagement with the partner's perspective \citep{slovic2002affect}. Their prevalence may therefore reflect not their decision-closing power but their role in sustaining the interactional substrate on which decisions are built.

By contrast, rule-based heuristics such as Attribute Elimination carry a higher cognitive cost: they require the speaker to articulate explicit criteria, which demands more effortful processing \citep{payne1993adaptive}. They may also carry a social cost, as eliminating an option can implicitly reject a partner's suggestion. These dual costs help explain why such strategies are rarely initiated spontaneously, despite their effectiveness. When they do appear, they function as convergence mechanisms that move the conversation from open exploration to closure.

This pattern suggests that conversational decision-making proceeds through two functionally distinct phases: an \textit{exploration phase} dominated by low-cost, socially facilitative heuristics (affect, availability) that expand and maintain the decision space, and an \textit{exploitation phase} in which higher-cost, rule-based heuristics narrow the options and drive toward resolution. This two-phase structure parallels the exploration--exploitation trade-off described in computational models of decision-making \citep{cohen2007should}, but here it emerges from the interactional dynamics of conversation rather than from individual optimization. Future work could examine whether the timing of the transition between phases predicts decision success.

\subsubsection{Satisficing as a Conversational Achievement}

The predominance of satisficing over maximizing in our data is consistent with decades of work on bounded rationality \citep{simon1955behavioral, simon1956rational}, but our findings suggest that in conversational contexts, satisficing is not merely a default mode of individual reasoning but a joint accomplishment enabled by specific interactional mechanisms.

Choice Overload Mitigation (COM) and implicit agreement together scaffold the satisficing process. COM constrains the option space---typically through binary contrasts---so that a ``good enough'' threshold can be reached more quickly. Implicit agreement provides a low-cost pathway to acceptance without the exhaustive justification that maximizing would demand. In this sense, satisficing in conversation is a property of the dialogue system, not only of individual cognition.

\subsubsection{Internal Knowledge and Ecological Rationality}

The predominance of internal knowledge over external information retrieval supports the ecological rationality framework \citep{gigerenzer1999simple}, which argues that heuristic strategies are adapted to the structure of the decision environment. In everyday, low-stakes domains such as food and travel, the relevant information is often already available in memory---past experiences, personal preferences, situational constraints---making external search unnecessary or even counterproductive by adding cognitive load without proportionate benefit.

This finding extends the ecological rationality account from individual decision-making to conversational settings. In dialogue, the reliance on internal knowledge takes a specific form: participants retrieve and share personal experiences, articulate latent preferences, and use affective reactions as information. The conversational process thus functions less as a joint information search and more as a mutual preference clarification process.

\subsection{Transferability to Human--AI Interaction}

Applying findings from human--human conversation to human--AI design requires addressing a structural difference: human--human decision-making often involves joint commitment, where both parties must agree to act on the outcome, whereas human--AI interaction is typically framed as individual decision support. We address this at two levels.
 
First, the core cognitive patterns identified in this study---satisficing, reliance on internal knowledge, the frequency--efficiency mismatch, and choice overload mitigation---are properties of individual bounded rationality that surface in conversation but are not dependent on joint commitment. These patterns arise from the decision-maker's own cognitive constraints and would be expected to operate regardless of whether the conversational partner is human or artificial.
 
Second, the social patterns in our data are likely to appear in human--AI interaction as well, both because users apply social heuristics to machines exhibiting conversational cues \citep{nass2000machines, gambino2020building} and because AI systems themselves increasingly occupy relational roles---from neutral information provider to opinionated advisor to persistent conversational partner with accumulated user knowledge. The role a conversational AI occupies modulates which patterns from our codebook apply most directly, but across this range, strategies such as implicit rejection, conversational deferral, and suspension remain relevant. A design-critical implication is that the absence of explicit negation should not be interpreted as confirmation, precisely because users bring social cognition to machine interlocutors.
 
Rather than treating human--human and human--AI decision-making as categorically distinct, we suggest that they occupy a shared spectrum of conversational decision structures, differentiated by the degree of commitment, social obligation, and role asymmetry between participants. Our codebook is designed as a flexible analytical resource that can be selectively applied across this spectrum.

\subsection{Design Implications}

Building on the theoretical analysis and transferability arguments above, we propose the following design guidelines grounded in the observed cognitive and interactional patterns.

\textbf{1. Memory-Based Preference Elicitation.} Given the predominance of internal knowledge and the interpretation of everyday decisions as preference clarification rather than information search, systems should minimize the cognitive cost of external retrieval. Instead of querying attribute parameters (e.g., price, location), the dialogue manager should employ associative prompting---triggering episodic memory to surface latent preferences (e.g., ``Do you recall a similar experience you enjoyed?''). This shifts the system function from database querying to memory support.
 
\textbf{2. Operationalizing Implicit Rejection.} Standard intent recognition models often classify non-response or topic shifts as ``neutral'' or ``continuation.'' However, given the prevalence of implicit strategies and the rarity of explicit rejection in our data, and given that the CASA literature predicts users will employ these same face-saving strategies with AI interlocutors \citep{nass2000machines}, dialogue policies should be calibrated to interpret ambiguous user turns (e.g., hesitation, unrelated inquiries) as potential negative feedback signals, triggering exploration of alternative branches rather than persisting with the current candidate.
 
\textbf{3. Structured Pairwise Presentation.} To address choice overload, systems should avoid presenting unstructured recommendation lists. Instead, the interface should adopt a pairwise comparison framework, presenting options in pairs (e.g., A vs.\ B). This aligns with the conversational satisficing patterns observed in our data, reducing the cognitive load of global optimization to a simpler task of local preference judgment \citep{iyengar2000choice}.
 
\textbf{4. Phase-Adaptive Heuristic Support.} The two-phase structure identified in our theoretical analysis provides a concrete policy for conversational agents. During the \textit{exploration phase}, the system should process and mirror affective and availability cues to maintain engagement and build shared context. As the conversation progresses to the \textit{exploitation phase}---signaled by turn count, heuristic cycling, or user signs of decision fatigue---the system should actively supply the rule-based mechanisms that users rarely initiate spontaneously, introducing constraint-based proposals (e.g., ``Excluding options over distance $X$'') to drive toward closure. This phase-adaptive strategy addresses the frequency--efficiency mismatch by providing the right kind of support at each stage rather than defaulting to a single mode.

\section{Limitations and Future Work}

This study is limited by the scope of its data and setting. The study focuses on Korean everyday decision scenarios, which may limit generalizability to other cultural contexts or domains (e.g., medicine, finance). Second, the conversations are text-based and may differ from multimodal or face-to-face interactions. Third, our LLM-assisted coding pipeline, while calibrated against human coders, may still contain residual annotation errors. Future work can extend this analysis to additional domains and languages, investigate sequential patterns (e.g., how specific strategies unfold turn by turn), and embed these findings into the design and evaluation of conversational AI agents for decision support.

\section{Conclusion}

We presented a conversation-analytic study of everyday decision-making dialogue, using Korean conversation datasets to develop and apply a codebook of decision strategies, conversational moves, and heuristics. Our findings show that satisficing, internal knowledge, affective judgments, and choice-overload mitigation are central to how people talk their way toward decisions, and that specific heuristic patterns are associated with whether conversations reach resolution. These insights provide a grounded foundation for designing conversational AI that complements bounded rationality and human heuristics, supporting everyday decision-making in a way that aligns with how people actually think and converse.

\section{Acknowledgments}
This work was partly supported by Institute of Information \& Communications Technology Planning \& Evaluation (IITP) grant funded by the Korea government (MSIT) [NO.RS-2021-II211343, Artificial Intelligence Graduate School Program (Seoul National University)], SNU-Global Excellence Research Center establishment project, and Basic Science Research Program through the National Research Foundation of Korea (NRF) funded by the Ministry of Education (No. RS-2025-25421701).

\printbibliography

\end{document}